\documentclass[twocolumn,prl,aps,superscriptaddress,showpacs]{revtex4}
\usepackage{amsmath}
\usepackage{graphicx}
\usepackage{txfonts}
\usepackage{hyperref}
\usepackage{amsmath}
\usepackage{amsfonts}
\usepackage{amssymb}
\usepackage{graphicx}%
\providecommand{\U}[1]{\protect\rule{.1in}{.1in}}
\begin{document}
\title{Quantum State Preparation with Universal Gate Decompositions}
\author{Martin Plesch}
\affiliation{Faculty of Physics, University of Vienna, Vienna,
Austria} \affiliation{Faculty of Informatics, Masaryk University,
Brno, Czech Republic}
 \affiliation{Institute of Physics,
Slovak Academy of Sciences, Bratislava, Slovakia}
\author{\v{C}aslav Brukner}
\affiliation{Faculty of Physics, University of Vienna, Vienna, Austria} \affiliation{Institute of Quantum Optics and Quantum
Information (IQOQI), Austrian Academy of Sciences, Vienna, Austria}

\date{17 January 2011}

\begin{abstract}
In quantum computation every unitary operation can be decomposed
into quantum circuits -- a series of single-qubit rotations and a
single type entangling two-qubit gates, such as Controlled-NOT
(C-NOT) gates. Two measures are important when judging the
complexity of the circuit: the total number of C-NOT gates needed
to implement it and the depth of the circuit, measured by the
minimal number of computation steps needed to perform it. Here we
give an explicit and simple quantum circuit scheme for preparation
of arbitrary quantum states, which can directly utilize any
decomposition scheme for arbitrary full quantum gates, thus
connecting the two problems. Our circuit reduces the depth of best
currently known circuit by a factor of $2$. It also reduces the
total number of C-NOT gates from $2^n$ to $\frac{23}{24}2^n$ in
the leading order for even number of qubits. Specifically, the
scheme allows to decrease the upper bound from $11$ C-NOT gates to
$9$ and the depth from $11$ to $5$ steps for four qubits. Our
results are expected to help in designing and building small-scale
quantum circuits using present technologies.

\end{abstract}


\maketitle

%%%%%%%%%%%%%%%%%%%%%%%%%%%%%%%%%%%%%%%%%%%%%%%%%%%%%%%%%%

\section{Introduction}

Quantum information and computation theory (\cite{Nielsen} and
references therein) is receiving attention in last decades due to
its possibility to outperform information processing based on the
classical physics in the areas of secure communication~\cite{BB84}
or efficient implementation of certain computation tasks, e.g.,
prime number factorization~\cite{Shor}.

Similarly to classical computation, every quantum computation,
represented as a unitary operation performed on a desired state of
qubits, can be decomposed into small operation blocks, where only
a subset of qubits is changed non-trivially. Whereas one-qubit
operations cannot be composed to a general unitary operation, as
they never change the degree of entanglement within the state, a
single type of two-qubit operation (for example,
C-NOT~\cite{footnote}) in combination with arbitrary one-qubit
rotations suffices \cite{Kniznica}.

%One of the main difficulties that prevents a broad practical application of quantum computers is impossibility to perform quantum
%operations within a precision that is needed to perform quantum error correcting schemes.

The complexity of quantum circuits is usually measured in the
number of C-NOT gates needed to perform the desired unitary
operation. The reason to count the number of two-qubit gates is
mainly experimental since their realization is much more demanding
and introduces more imperfections than the realization of
one-qubit gates. Adding every new C-NOT to the circuit increases
its overall imperfection. This constitutes the main obstacle
preventing realization of quantum computation within sufficient
precision. It is therefore crucial to design circuits with the
least possible number of entangling gates.

In general, an exponential number of C-NOT gates with respect to
the number of qubits involved is needed to implement a general
unitary operation. This can be seen by simple counting of
parameters of an $n$-qubit unitary operation. Several attempts
have been made to optimize the number of gates needed for general
operations~\cite{Cosin,twoqubits,Sedlak,Saeedi,State2,Vartiainen,Vartiainen2,Bounds,Nakajima,Drury}.

In situations where the input for a quantum computer or a quantum
communication protocol is a known quantum state, we are not
interested to perform a completely defined unitary transformation.
Instead, we aim only to prepare a given state $| \phi \rangle$,
i.e. to perform a transformation from an initial state $| \psi
\rangle$ to a different target state: $| \psi \rangle
\rightarrow\left\vert \phi\right\rangle $, where a whole class of
unitaries $U$ fulfills the condition $U\left\vert
\psi\right\rangle =\left\vert \phi\right\rangle $.

It is known that one needs an exponential number of C-NOT gates to
prepare a generic quantum state, i.e., in the leading order this
number is $N_{C-NOT}=c \cdot 2^n$ , where $c$ is a pre-factor and
$n$ is the number of qubits. Any optimization can only decrease
the pre-factor but cannot beat the exponential dependence. The
best known result so far is $c=1$~\cite{State2}. Here we give an
explicit quantum circuit reducing the pre-factor to
$c=\frac{23}{24}$ for $n$ even. Specifically, using our scheme we
decrease the known upper bound from $11$ C-NOT gates to $9$ for
four qubits and from $57$ C-NOT gates down to $46$ for six qubits,
keeping the existing bound of $26$ C-NOT gates for five qubits.
The lower bounds are $6$, $13$ and $29$ C-NOT gates respectively
(see below).

The reduction of the overall number of C-NOT gates might be,
however, not the only aim of the optimization procedure. Searching
for efficient algorithms, the depth of the quantum circuit, i.e.
the minimal number of computation steps required for accomplishing
the computation, is crucial~\cite{footnote2}. In a general case,
the depth might be as high as the overall number of C-NOT gates,
not allowing to perform more than one gate in parallel as is the
case  in Ref.~\cite{State2}. In our scheme the depth is at most
half of the number of C-NOT gates, i.e. at least two gates can be
implemented in parallel in every step.

\section{Lower bounds}

A general $n$-qubit pure state is fully described by $2^{n+1}-2$
real parameters. During the preparation process, these parameters
are introduced sequently by performing single-qubit rotations
(each rotation introduces three Euler angles) along with C-NOT
gates. C-NOT gates as such do not introduce any parameters, but
they are a kind of barriers that separate one-qubit rotations such
that they cannot merge into a resulting single rotation for each
qubit. Naively, one could expect that every C-NOT gate can be
accompanied with two one-qubit operations -- one for the control
and one for the target qubit -- applied after every C-NOT gate.
Due to existing identities~\cite{Bounds}, however, only four real
parameters can be introduced with one C-NOT gate. This can be
understood as follows: Rotation about $z$ axis applied on the
control qubit commutes with the C-NOT gate. Similarly, rotation
about $x$ axis applied on the target qubit commutes with the C-NOT
gate. In this way, the two types of rotations can be commuted
backward through the C-NOT gate and combined with the rotations
applied after the previous C-NOT gates acting on the respective
qubits. Thus for every C-NOT we can implement four real parameters
of the desired state.

Further parameters can be added by local unitary transformations
on qubits in the beginning of the process. Trivially, one would
expect to introduce three real parameters per qubit, corresponding
to three Euler angles. However, this is not the case. By starting
in a specific product state (e.g.
$\left\vert0\right\rangle^{\otimes N}$), we may only rotate every
single qubit into a given direction, what gives us two parameters
per qubit. The third, missing parameter, is just a phase on every
qubit, which sums up through all qubits and influences only the
global phase. Therefore on $n$ qubits, with $k$ C-NOT gates, we
may introduce altogether up to $4k+2n$ real parameters. This gives
a lower bound on the number of C-NOT gates needed to prepare a
state: $6$ for four qubits, $13$ for five qubits and $29$ for six
qubits. For large numbers of $n$ we get a lower bound on the
number of C-NOT gates $k=\frac{1}{2}2^{n}$ in the leading order.

The lower limit for the depth of the circuit also grows
exponentially with the number of qubits, with a linear correction.
This can be seen from the fact that in one computation step no
more than $\frac{n}{2}$ C-NOT gates can be performed. The only
possible optimization for the depth is also the reduction of the
pre-factor with up to a linear correction, with the lower bound
$\frac{2^n}{n}$.

%It is known that one needs an exponential number of C-NOT gates to
%prepare a generic quantum state, i.e. this number is $N_{CNOT}
%\propto c \cdot 2^n$, where $c$ is a pre-factor and $n$ is the
%number of qubits.

%For odd number of qubits our scheme does not bring a reduction in
%the overall number of CNOT gates, however its simplicity and high
%degree of possible parallelism in performing different C-NOT gates
%in one time instance make our scheme interesting also for odd
%number of qubits.

%In this work, we show how to achieve the pre-factor
%$\frac{23}{24}$ for even number of qubits, but further
%optimization might be possible. For four qubits the lower limit is
%$6$ CNOT gates and for six qubits $29$ CNOT gates.

\section{Four qubits}

The Hilbert space of four qubits can be factorized into two
parts, where each part is associated to two qubits. An
arbitrary pure state $|\Psi\rangle$ of four qubits can then
be expressed using the (standard) Schmidt decomposition as
\begin{equation}
\left\vert \Psi\right\rangle =\sum_{i=1}^{4}\alpha_{i}\left\vert
\psi\right\rangle _{i}\left\vert \phi\right\rangle _{i}. \label{Schmidt}%
\end{equation}
Here $\left\vert \psi\right\rangle _{i}$, $i=1,...,4$, are
four normalized orthogonal states of the first two qubits
and similarly $\left\vert \phi\right\rangle _{i}$ are four
normalized orthogonal states of the second two qubits. The
states are given with a nontrivial global phase. The
coefficients $\alpha_{i}$ are real and positive and they
obey $\sum_{i=1}^{4}\alpha_{i}^{2}=1$. Without the loss of
generality we can rewrite the decomposition~(\ref{Schmidt})
in such a way that $\left\vert \psi\right\rangle _{i}$ and
$\left\vert \phi\right\rangle _{i}$ will be defined only up
to a global phase. Their relative phases (with respect to
different $i$'s) will then be included in the generalized
coefficients $\alpha_{i}$, which become complex. As we are
interested in $\left\vert \Psi\right\rangle $ up to its
global phase, we can make the choice of having $\alpha_{1}$
real positive.

The pure state $\left\vert \Psi\right\rangle $ is specified by
$2^{5}-2=30$ real parameters. The four states $\left\vert
\psi\right\rangle _{i}$ are specified by $6$, $4$, $2$ and $0$
parameters (due to orthogonality condition), and so are the four
states $\left\vert \phi\right\rangle _{i}$. The four coefficients
$\alpha_{i}$ require $6$ independent real parameters to be
determined due to normalization condition and the choice of the
global phase. This gives altogether $30$ parameters, as expected.

\subsection{Phase 1}

To prepare the state $\left\vert \Psi\right\rangle $
starting from the initial state $\left\vert
0000\right\rangle $, we first generate the state with the
generalized (complex) Schmidt coefficients on the first two
qubits:
\begin{equation}
\left\vert 0000\right\rangle \rightarrow\left(  \alpha_{1}\left\vert
00\right\rangle +\alpha_{2}\left\vert 01\right\rangle +\alpha_{3}\left\vert
10\right\rangle +\alpha_{4}\left\vert 11\right\rangle \right)  \left\vert
00\right\rangle. \label{Step1}%
\end{equation}
This operation does not define a unitary operation
completely, but is a state-preparation operation on two
qubits (starting from a known state $\left\vert
00\right\rangle$ we end in a state specified by the
generalized Schmidt decomposition coefficients). Therefore,
as shown in Ref.~\cite{Cosin}, it can be realized by one
C-NOT operation in combination with suitable one-qubit
rotations.

\subsection{Phase 2}

We perform two C-NOT operations, one with the control on the first
qubit and the target on the third qubit and the other one with the
control on the second qubit and the target on the fourth qubit. In
such a way we can ``copy'' the basis states of the first two
qubits onto the respective states of the second two qubits. In
this way we obtain a state of four qubits, which has the same
Schmidt decomposition coefficients as the target
state~(\ref{Schmidt}).
\begin{align}
&  \left(  \alpha_{1}\left\vert 00\right\rangle +\alpha_{2}\left\vert
01\right\rangle +\alpha_{3}\left\vert 10\right\rangle +\alpha_{4}\left\vert
11\right\rangle \right)  \left\vert 00\right\rangle \rightarrow\label{Step2} \\
&  \rightarrow\left(  \alpha_{1}\left\vert 00\right\rangle \left\vert
00\right\rangle +\alpha_{2}\left\vert 01\right\rangle \left\vert
01\right\rangle +\alpha_{3}\left\vert 10\right\rangle \left\vert
10\right\rangle +\alpha_{4}\left\vert 11\right\rangle \left\vert
11\right\rangle \right)  .\nonumber
\end{align}
For this phase we obviously only need two C-NOT operations;
one-qubit rotations are not necessary.

\subsection{Phase 3}

Keeping the Schmidt decomposition form we apply the unitary
operation that transforms the basis states of the first two qubits
into the four states $\left\vert \psi\right\rangle _{i}$. We
obtain:
\begin{align}
\left\vert 00\right\rangle    \rightarrow\left\vert
\psi\right\rangle _{1} && \left\vert 01\right\rangle
\rightarrow\left\vert \psi\right\rangle
_{2}\label{Step3}\\
\left\vert 10\right\rangle   \rightarrow\left\vert
\psi\right\rangle _{3} && \left\vert 11\right\rangle
\rightarrow\left\vert \psi\right\rangle _{4}.\nonumber
\end{align}
As for any two-qubit unitary operation we do not need more than 3 C-NOT gates
\cite{Cosin}.

\subsection{Phase 4}

In the final phase of the circuit we perform a unitary
operation on the third and fourth qubit in order to
transform their computational basis states into the Schmidt
basis states of Eq.~(\ref{Schmidt}):
\begin{align}
\left\vert 00\right\rangle  \rightarrow\left\vert
\phi\right\rangle _{1} && \left\vert 01\right\rangle
\rightarrow\left\vert \phi\right\rangle
_{2}\label{Step4}\\
\left\vert 10\right\rangle  \rightarrow\left\vert
\phi\right\rangle _{3} &&\left\vert 11\right\rangle
\rightarrow\left\vert \phi\right\rangle _{4}.\nonumber
\end{align}
Similarly to the previous phase, we again use 3 C-NOT operations.
We conclude that altogether we have used $1+2+3+3=9$ C-NOT gates
for the entire quantum state preparation circuit, which is less
than the best result of $11$ C-NOT gates, which can be deduced
from \cite{State2}. However, it stays above the minimum of $6$
gates obtained from parameter counting.

The depth of the circuit is $5$, where the second phase can be
done in one computation step and the third and fourth phases can
be done in parallel in three computation steps. This is less than
half of the result of Ref.~\cite{State2} and is optimal for $9$
C-NOT gates. The theoretical minimal depth is $3$, deduced from
the fact that at least $6$ C-NOT gates are needed and no more than
two can be performed in one step.
%TCIMACRO{\FRAME{ftbpFU}{2.7959in}{1.516in}{0pt}{\Qcb{Gate sequence for
%preparation of an arbitrary four qubit state. Between C-NOT gates individual
%one-qubit rotations need to be applied. }}{\Qlb{figure1}}{4qubits.png}%
%width 2.7959in;  height 1.516in;  depth 0pt;  original-width 2.7536in;
%original-height 1.4797in;  cropleft "0";  croptop "1";  cropright "1";
%cropbottom "0";  filename '4qubits.png';file-properties "XNPEU";}}}%
%BeginExpansion

\begin{figure}
[ptb]
\begin{center}
\includegraphics[
%%natheight=1.479700in,
%%natwidth=2.753600in,
%%height=1.516in,
width=2.7536in ] {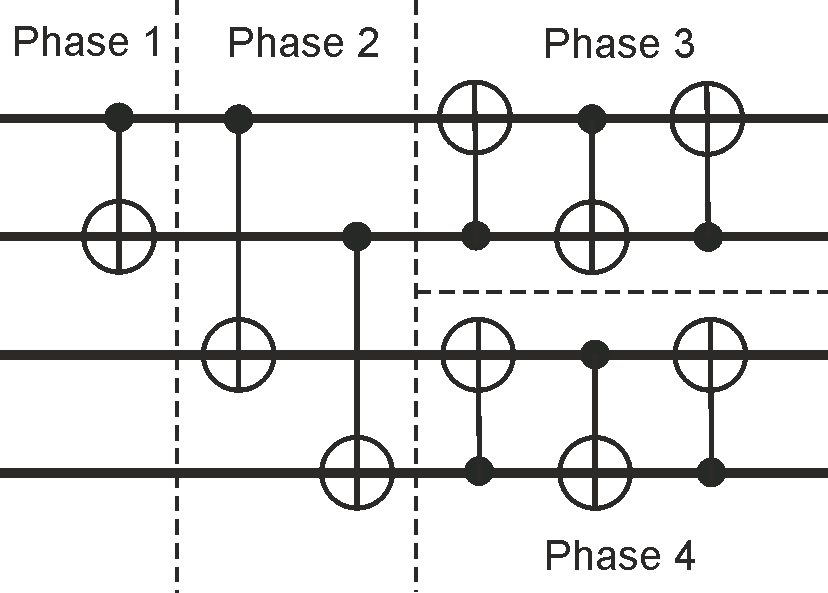}
 \caption{Gate sequence for preparation of an arbitrary four-qubit state.
Individual one-qubit rotations (not depicted) need to be
applied between C-NOT gates. The four individual phases are
described in the text. The two C-NOT gates in phase 2, as
well as in the phases 3 and 4 can be performed in parallel,
as they address different qubits. Altogether one needs 9
C-NOT gates, 4 pairs of which can be performed in
parallel.} \label{figure1}
\end{center}
\end{figure}

%EndExpansion

\section{Five qubits}

To illustrate our state preparation procedure for the case
of odd number of qubits -- where the entire Hilbert space
cannot be factorized into Hilbert spaces of equal
dimensions -- we give an example for five qubits. We first
factorize the Hilbert space into two parts, with one part
associated with two qubits and the other one with three
qubits. The Schmidt decomposition of an arbitrary
five-qubit state with respect to such Hilbert space
factorization has almost the same structure as in the case
of four qubits in Eq.~(\ref{Schmidt}). One has
\begin{equation}
\left\vert \Psi\right\rangle =\sum_{i=1}^{4}\alpha_{i}\left\vert
\psi\right\rangle _{i}\left\vert \phi\right\rangle _{i}. \label{5Schmidt}%
\end{equation}
Again, the summation goes at most over four terms and the
only difference is that states $\left\vert
\phi\right\rangle _{i}$, $i=1,...,4$, are now three-qubit
states. We again choose to include the relative phase of
the states into the coefficients $\alpha_{i}$ and proceed
with the phases one to three in the same way as for four
qubits. The only difference is in the fourth phase, where
we perform a three-qubit unitary operation:
\begin{align}
\left\vert 00\right\rangle \left\vert 0\right\rangle
\rightarrow\left\vert \phi\right\rangle _{1} && \left\vert
01\right\rangle \left\vert 0\right\rangle   \rightarrow\left\vert
\phi\right\rangle _{2}\label{5Step4}\\
 \left\vert 10\right\rangle
\left\vert 0\right\rangle    \rightarrow\left\vert
\phi\right\rangle _{3} && \left\vert 11\right\rangle \left\vert
0\right\rangle  \rightarrow\left\vert \phi\right\rangle
_{4}.\nonumber
\end{align}
Such unitary can be implemented by no more than $20$ C-NOT
gates~\cite{Cosin}. Moreover, this unitary is not completely
defined (the third qubit is initially exclusively in the state
$\left\vert 0\right\rangle $) and thus further reduction of the
number of C-NOT gates might be possible. Even without such
optimizations, our state preparation procedure for five qubits
requires $1+2+3+20=26$ C-NOT gates, which achieves result of
Ref.~\cite{State2}. The lover limit of $13$ C-NOT gates suggests
that further optimization is possible.

The depth of the procedure is $22$ computation steps, with one
step for phase one, one step for phase two and $20$ steps for
performing the phases three and four in parallel. This is less
than the lowest known depth of $26$ of Ref.~\cite{State2}, but
more than the theoretical lower bound of $7$.

\section{General case}

We will now apply the main idea presented for four and five qubits
to the general case of $n$ qubits. We begin with factorization of
the Hilbert space of $n$ qubits into two parts of equal dimension
for $n$ even, so that each part is associated to $\frac{n}{2}$
qubits. For odd number of qubits we factorize the Hilbert space
into $\frac{n-1}{2}$ and $\frac{n+1}{2}$ qubits. On the first part
of the qubits we will prepare a state whose amplitudes in the
computational basis will be defined by the generalized Schmidt
coefficients. Then we will apply a set of C-NOT gates between the
qubits in the first and the second part. In the end we will
perform two unitary operations, one on the first part and one on
the second part of qubits. We will separately treat the case of
even and odd number of qubits.

\subsection{Even number of qubits}

We write the number of qubits as $n=2k$. The qubits are
divided into two parts, each containing $k$ qubits. With
respect to this division the Schmidt decomposition of an
arbitrary state of $n$ qubits has the following form
\begin{equation}
| \Psi\rangle =\sum_{i=1}^{2^{k}}\alpha_{i} \left\vert
\psi\right\rangle_{i}\left\vert \phi\right\rangle_{i}, \label{nSchmidt}%
\end{equation}
where both $\left\vert \psi\right\rangle _{i}$ and $\left\vert \phi
\right\rangle _{i}$ are normalized states of $k$ qubits and $\alpha_{i}$
are complex coefficients.

The initial state of qubits is assumed to be the product state
$\left\vert 0\right\rangle ^{\otimes2k}$ in which each qubit is in
state $|0\rangle$. On the first $k$ qubits we prepare a
superposition state whose amplitudes are the Schmidt coefficients
in the computational basis:
\begin{equation}
\left\vert 0\right\rangle ^{\otimes2k}\rightarrow\left(  \sum_{i=1}^{2^{k}%
}\alpha_{i}\left\vert i\right\rangle \right)  \left\vert 0\right\rangle
^{\otimes k}. \label{nStep1}%
\end{equation}
The sequence of 0's and 1's in the computational basis
states
$\{|00...0\rangle,|00...1\rangle,...,|11...1\rangle\}$
represents the binary encoding of the index $i$ in the
states $\left\vert i\right\rangle$: Qubits in state
$\left\vert 1\right\rangle $ stand exactly on those
positions where there is a $1$ in the binary notation of
$i$. All other qubits are in the state $\left\vert
0\right\rangle $. To prepare a state on $k$ qubits as
required in Eq.~(\ref{nStep1}), we can utilize the existing
bound from Ref.~\cite{State2}, which allows us to prepare
it with help of $2^{k}-k-1$ C-NOT gates. We will later
return to the discussion about further optimization
possibilities of this particular phase.

In the second phase, we perform $k$ C-NOT gates with qubits
$j$ as the control and qubits $j+k$ as the target for $j$
running from $1$ to $k$. This will bring us to the desired
Schmidt form of our state
\begin{equation}
\left(  \sum_{i=1}^{2^{k}}\alpha_{i}\left\vert i\right\rangle \right)
\left\vert 0\right\rangle ^{\otimes k}\rightarrow\sum_{i=1}^{2^{k}}\alpha
_{i}\left\vert i\right\rangle \left\vert i\right\rangle . \label{nStep2}%
\end{equation}

The phases three and four are $k$-qubit unitary operations
performed on the first and second half of qubits respectively. We
obtain
\begin{align}
\sum_{i=1}^{2^{k}}\alpha_{i}\left\vert i\right\rangle \left\vert
i\right\rangle  &  \rightarrow\sum_{i=1}^{2^{k}}\alpha_{i}\left\vert
\psi\right\rangle _{i}\left\vert i\right\rangle \label{nStep3}\\
&  \rightarrow\sum_{i=1}^{2^{k}}\alpha_{i}\left\vert \psi\right\rangle
_{i}\left\vert \phi\right\rangle _{i}, \label{nStep4}%
\end{align}
which is the aimed target state~(\ref{nSchmidt}). Every unitary
operation acting on $k$ qubits can be performed by $\frac{23}{48}2^{2k}%
-\frac{3}{2}2^{k}+\frac{4}{3}$ C-NOT gates \cite{Cosin}. We thus
need altogether $2^{k}-k-1+k+\frac{23}{24}2^{2k}-\frac
{3}{2}2^{k+1}+\frac{8}{3}$ C-NOT gates. This number is bounded
from above by its leading term in $k$. Taking $n=2k$ we obtain
\begin{equation}
N_{CNOT}^{even}<\frac{23}{24}2^{n}. \label{nEven}%
\end{equation}
This is the new lowest number of C-NOT gates needed for
construction of a universal circuit for preparation of an
arbitrary state.

In the first phase~(\ref{nStep1}) of the procedure given above we
used a method for state preparation, which requires more
entangling gates than our method. Naturally, we can use our result
recursively to obtain a slightly lower number of C-NOT operations
needed to prepare the state of the first $k$ qubits. However, this
part of the process does not contribute to the leading order of
the number of C-NOT gates needed for preparation as calculated in
Eq.~(\ref{nEven}). The first phase contributes only with the order
of $2^\frac{n}{2}$, whereas the phases three and four contribute
with the order of $2^n$.

The depth of the circuit is, in the leading order, given by the
depth of the phases three and four, which is $\frac{23}{48}2^{n}$,
less than the best previous result of $2^{n}$, but weaker than the
theoretical limit of $\frac{2^{n}}{n}$.

\subsection{Odd number of qubits}

We express the number of qubits as $n=2k+1$. The first
three phases, as described by
equations~(\ref{nStep1},\ref{nStep2},\ref{nStep3}) of the
procedure remain exactly the same as for the case of even
number of qubits. In the phase four~(\ref{nStep4}) we
perform a unitary operation on $k+1$ instead of $k$ qubits.
Summing up contributions from all four phases we obtain the
overall number of C-NOT gates required to be
$2^{k}-k-1+k+\frac{23}{48}2^{2k}-\frac
{3}{2}2^{k}+\frac{4}{3}+\frac {23}{48}2^{2k+2}-\frac
{3}{2}2^{k+1}+\frac{4}{3}$. Similarly to the previous case
the leading order of this sum bounds the number of the
C-NOT gates from above. It can be simplified to
%\begin{equation}
$N_{CNOT}^{odd}<\frac{115}{96}2^{n}$.
%\end{equation}
This result is weaker then the bound~(\ref{nEven}) for even number
of qubits. However, further optimizations are possible since in
phase four the operation required is not a completely defined
unitary and one does not necessarily need the whole number of
C-NOT gates as required for a general unitary rotation on $k+1$
qubits. Moreover, even in this case the depth of the circuit
bounded by $\frac{115}{192}2^{n}$ is smaller than the best known
result.

\section{Conclusions}

We give an explicit and efficient circuit for preparation
of arbitrary states of $n$ qubits using a gate library
consisting of a single two-qubit gate (C-NOT) and one-qubit
rotations. For even number of qubits we have slightly
reduced the previously known upper bound on the number of
C-NOT gates needed. For the special case of four qubits our
scheme requires only $9$ C-NOT gates (compared to $11$
previously known), which should be within the scope of near
future quantum technology.

Our quantum state preparation scheme provides also a lower
computational depth than the previously known results. It can be
divided into four phases, where the last two can be performed in
parallel, which leads to roughly half of computational steps
comparing to the previous results. This opens further optimization
possibilities for experimental implementation of the state
preparation. Our results can help in designing and building
small-scale quantum circuits using present technologies (see,
e.g., Refs.~\cite{Experiment,Experiment2}).

Our procedure introduces a conceptually simple utilization of
efficient decomposition of arbitrary quantum gates for the problem
of state preparation. In fact, the efficiency of our procedure is
based on the best results for gate decompositions. If better
results will be obtained in future, they will directly lead to
lowering of our bounds. Moreover, this utilization itself is very
efficient: a circuit for gate decomposition reaching the lower
bound of $4^{(n-2)}$ \cite{Bounds} C-NOT gates in leading order
would lead to state preparation with $2^{(n-1)}$ C-NOT gates,
reaching the lower bound in the leading order as well.

Using our scheme one can also efficiently apply operations
that transform any given state $\left\vert
\psi\right\rangle $ of $n$ qubits to any other given state
$\left\vert \phi\right\rangle $. We first run the
preparation procedure for $\left\vert \psi\right\rangle $
in the reversed order, which results in the state
$\left\vert 0\right\rangle ^{\otimes n}$. Then, we continue
with preparing the aimed state $\left\vert
\phi\right\rangle $. The number of C-NOT gates needed to
perform this composite transformation is just double the
number needed to prepare an arbitrary state from
$\left\vert 0\right\rangle ^{\otimes n}$. However, the
depth of the complete circuit is even less than double, as
the last phase of the reversed process and the first phase
of the preparation process can run on distinct qubits and
therefore be performed in parallel.

%%%%%%%%%%%%%%%%%%%%%%%%%%%%%%%%%%%%%%%%%%%%%%%%%%%%%%%%%%%%%%%%%%%%%%%%%%%%

\section{Acknowledgments}

This work was supported by the Action Austria -- Slovakia,
SoMoPro project SIGA 862, the Austrian Science Foundation
FWF within Projects No. P19570-N16, SFB and CoQuS No.
W1210-N16, and the European Commission Project QESSENCE.
The collaboration is a part of \"OAD/APVV SK-AT-0015-10
project.
%%%%%%%%%%%%%%%%%%%%%%%%%%%%%%%%%%%%%%%%%%%%%%%%%%%%%%%%%%%%%%%%%%%%%%%%%%%%%
%%%%%%%%%%%%%%%%%%%%%%%%%%%%%%%%%%%%%%%%%%%%%%%%%%%%%%%%%%%%%%%%%%%%%%%%%%%%%


\begin{thebibliography}{99}
\bibitem{footnote} C-NOT, or Controlled-NOT operation reverts the state of the target qubit if and only if the control qubit
is in the state $|1\rangle$. It can be described in the computation basis as $|00\rangle\rightarrow|00\rangle$,
$|01\rangle\rightarrow|01\rangle$, $|10\rangle\rightarrow|11\rangle$ and $|11\rangle\rightarrow|10\rangle$,  where e.g. $|00\rangle$
denotes a product state in which both qubits are in $|0\rangle$ state.
\bibitem{footnote2} As a "computation step" only the performance of
C-NOT gates is considered. Several C-NOT gates performed in
parallel are considered as one step.
\end{thebibliography}
\end{document}